\documentclass[11pt]{article}

\pdfoutput=1
\usepackage{amssymb,amsmath}
\usepackage[dvipsnames]{xcolor}
\usepackage{latexsym}
\usepackage{mathrsfs}
\usepackage{verbatim,graphicx}
\usepackage{bm}
\usepackage{cite}
\usepackage[%
             colorlinks=true,urlcolor=MidnightBlue,linkcolor=MidnightBlue,citecolor=OliveGreen,
             pdfpagelabels=true,hypertexnames=true,
            plainpages=false,naturalnames=false,
             ]{hyperref}

\def\bea{\begin{eqnarray}}
\def\eea{\end{eqnarray}}
\def\be{\begin{equation}}
\def\ee{\end{equation}}
\def\ba{\begin{array}}
\def\ea{\end{array}}

\def\nn{\nonumber}

\begin{document}

\setlength\arraycolsep{2pt}

\setcounter{page}{1}

\begin{titlepage}

\rightline{\footnotesize{APCTP-Pre2014-009}} \vspace{-0.2cm}
\rightline{\footnotesize{KCL-PH-TH/2014-12}} \vspace{-0.2cm}

\begin{center}

\vskip 1.0 cm

{\LARGE  \bf On degenerate models of cosmic inflation}

\vskip 1.0cm

{\large
Rhiannon Gwyn$^{a}$, Gonzalo A. Palma$^{b}$, Mairi Sakellariadou$^{c}$ \\ and Spyros Sypsas$^{d}$
}

\vskip 0.5cm

{\it
$^{a}$Max-Planck-Institut f\"ur Gravitationsphysik, Albert-Einstein-Institut\\
\mbox{Am M\"uhlenberg 1, D-14476 Potsdam, Germany}
\\
$^{b}$Physics Department, FCFM, Universidad de Chile \\ \mbox{Blanco Encalada 2008, Santiago, Chile}
\\
$^{c}$Department of Physics, King's College London \\ \mbox{Strand, London WC2R 2LS, U.K.}
\\
$^{d}$Asia Pacific Center for Theoretical Physics (APCTP)\\ \mbox{Pohang 790-784, Republic of Korea}
}

\vskip 1.5cm

\end{center}

\begin{abstract}
In this article we discuss the role of current and future CMB measurements in pinning down the model of inflation responsible for the generation of primordial curvature perturbations. By considering a parameterization of the effective field theory of inflation with a modified dispersion relation arising from heavy fields, we derive the dependence of cosmological observables on the scale of heavy physics $\Lambda_{\rm UV}$. Specifically, we show how the $f_{\rm NL}$ non-linearity parameters are related to the phase velocity of curvature perturbations at horizon exit, which is parameterized by $\Lambda_{\rm UV}$. {\sc Bicep2} and {\sc Planck} findings are shown to be consistent with a value $\Lambda_{\rm UV} \sim \Lambda_{\rm GUT}$. However, we find a degeneracy in the parameter space of inflationary models that can only be resolved with a detailed knowledge of the shape of the non-Gaussian bispectrum. 
\end{abstract}

\end{titlepage}

\newpage

\section{Introduction \& summary} \label{intro}

Cosmic inflation \cite{Guth:1980zm,Linde:1981mu,Albrecht:1982wi} successfully explains the origin of the primordial curvature perturbations needed to seed the observed large-scale structure of our universe and the cosmic microwave background anisotropies \cite{Mukhanov:1981xt}. Its key predictions consist of a nearly Gaussian distribution of curvature perturbations characterized by a slightly red-tilted power spectrum, and the existence of primordial tensor modes. Cosmological observations have constrained various quantities, including the amplitude and spectral index of the power spectrum and, more recently, the tensor-to-scalar ratio~\cite{Komatsu:2010fb, Ade:2013ydc, Ade:2014xna}, to a point where a large number of inflationary models have already been discarded. Despite this progress, it is clear that more data is required in order to gain insight into the nature of the fundamental theory hosting inflation. One of the most promising avenues for this is the study of the small departures from Gaussianity parameterized by the three-point correlation function (or bi-spectrum) of curvature perturbations~\cite{Linde:1996gt, Bartolo:2001cw, Bernardeau:2002jy, Maldacena:2002vr, Lyth:2002my,Seery:2005wm}. The amplitude and shape of this function are known to be sensitive to the self-interactions dictating the non-linear evolution of fluctuations, as well as to their interactions with other possible degrees of freedom relevant at the time of horizon exit~\cite{Bartolo:2004if}.\footnote{Despite this, many degeneracies remain; see e.g. \cite{Gwyn:2012pb,Gwyn:2012ey}.}

The recent development of the effective field theory (EFT) framework \cite{Creminelli:2006xe, Cheung:2007st, Weinberg:2008hq, Senatore:2010wk, Khosravi:2012qg} to analyze the evolution of perturbations during inflation has been especially useful for discussing the potential existence of non-Gaussianity \cite{Creminelli:2005hu, Senatore:2009gt}. Using general symmetry arguments on a Friedman-Lema\^itre-Robertson-Walker (FLRW) space-time, the authors of ref.~\cite{Cheung:2007st} were able to deduce the most general action describing curvature fluctuations generated by a single degree of freedom. This formulation has led to a model-independent parameterization of curvature modes' self-interactions, exploiting the existence of non-linear relations among field operators of different orders in perturbation theory. 
In its simplest version, and up to cubic order, the EFT of inflation may be written in terms of a Goldstone boson field $\pi (t,x)$ parametrizing fluctuations along the broken time translation symmetry direction of the background, often written as
\be
S_{\rm EFT} = M_{\rm Pl}^2 \int \! d^3xdt  \, a^3 \epsilon H^2 \bigg[  \frac{1}{c_s^2} \dot \pi^2 -   \frac{(\nabla \pi)^2}{a^2}  + ( c_s^{-2}  - 1 ) \bigg(   \dot \pi^2 -  \frac{(\nabla \pi)^2}{a^2}   \bigg) \dot \pi  +  \frac{2 \tilde c_3}{3 c_s^2} (  c_s^{-2}  - 1) \dot \pi^3   \bigg],  \quad \label{Seft-simple}
\ee
where $a = a(t)$ is the scale factor, $H = \dot a/a$ is the Hubble expansion rate, $\epsilon = - \dot H / H^2$ is the usual slow roll parameter (terms sub-leading in the slow-roll parameters are omitted for convenience), and $c_{\rm s}$ denotes the speed of sound at which the Goldstone mode propagates. This quantity may be expressed in terms of a mass scale $M_2$ used in the EFT expansion of ref.~\cite{Cheung:2007st} as
\be \label{sos}
c_{\rm s}^{-2} = 1 + \frac{2M_2^4}{|\dot H|M_{\rm Pl}^2},
\ee
and will have a central role in our discussion.
The other variable, $\tilde{c}_3$, corresponds to a dimensionless quantity parametrizing non-linear interactions, and satisfies $\tilde{c}_3\propto M_3^4/M_2^4$, where $M_3$ is the next to leading order mass parameter in the EFT expansion.
In this formulation, the standard curvature perturbation $\mathcal{R}$ is given in terms of the Goldstone boson by $\mathcal{R} = - H \pi$. The values of $c_{\rm s}$ and $\tilde c_3$ characterize the cubic interactions, and are determined by the model being described. For instance, in single-field canonical models these two parameters take the values $c_{\rm s}=1$ and $\tilde c_3=0$, and the interactions are found to be suppressed with respect to the slow-roll parameters. In more exotic models, such as DBI inflation or multi-field inflation, the value of $c_{\rm s}$ may vary in time, but with values restricted to be lower --- or even much lower --- than $1$. In general, one expects the dimensionless parameter $\tilde c_3$ to be of order $1 - c_{\rm s}^2$, which follows from naturalness arguments~\cite{Senatore:2009gt}. For instance, in the particular case of DBI inflation~\cite{Alishahiha:2004eh} one finds $\tilde c_3 =3(1-c_{\rm s}^2)/2$, in the case of two-scalar field canonical models with a heavy field one has $\tilde c_3 =3(1-c_{\rm s}^2)/4$~\cite{Achucarro:2012sm}, whereas in models with two or more heavy fields one finds the bound $\tilde c_3 \geq 3(1-c_{\rm s}^2)/4$~\cite{Cespedes:2013rda}.

A suppressed value for the speed of sound changes the wavelength at which perturbations freeze, and increases the self-coupling between curvature perturbations, leading to the following formulas for the amplitude of the power spectrum $\Delta_{\mathcal{R}}$, tensor-to-scalar ratio $r$, and $f_{\rm NL}$ parameters (characterizing non-Gaussianity):
\be
\Delta_{\mathcal{R}}  = \frac{1.3}{100} \frac{H^2}{M_{\rm Pl}^2 \epsilon c_{\rm s}}  , \qquad   r = 16 \epsilon c_{\rm s} ,  \qquad f_{\rm NL} \sim \frac{1}{c_{\rm s}^2} .  \label{predictions-EFT}
\ee
We see immediately that within this effective field theory parametrization $H$ is uniquely determined by $\Delta_{\mathcal{R}}$ and $r$ via
\be
H = 2.2  \sqrt{r \Delta_{\mathcal{R}}} M_{\rm Pl}, \label{H-r-Delta}
\ee
which implies, using recent observations~\cite{Ade:2013ydc, Ade:2014xna}, a preferred value of  $H \simeq 10^{14}$ GeV for the Hubble parameter during inflation. However, current observations cannot resolve the values of the slow roll parameter $\epsilon$ and the speed of sound $c_{\rm s}$. Determining these quantities requires better non-Gaussian constraints on the various $f_{\rm NL}$ parameters. The sensitivity of $f_{\rm NL}$ on $c_{\rm s}$ has turned the speed of sound into a powerful parametrization of models beyond the single-field canonical paradigm. Current searches of non-Gaussianity~\cite{Ade:2013ydc} constrain the speed of sound to lie in the range $0.02 \leq c_{\rm s} \leq 1$. 

More elaborate parameterizations of inflation are also possible within the EFT framework~\cite{Cheung:2007st}. For instance, it was argued on general grounds in ref.~\cite{Baumann:2011su} that, for short enough wavelengths of the curvature perturbations, the EFT could exhibit a non-trivial scaling of its field operators, enhanced by the broken time translation invariance of the background. For this to be possible, a new mass parameter needs to enter the EFT description, introducing a pivot scale at which this new scaling becomes operative. An example of such an EFT is obtained in the particular case where curvature perturbations interact with heavy scalar degrees of freedom, with masses $\Lambda_{\rm UV}$ such that $H \ll \Lambda_{\rm UV}$. In this type of scenario, if the speed of sound and the Hubble scale satisfy $c_{\rm s}^2 \Lambda_{\rm UV}  \ll  H \ll \Lambda_{\rm UV}$, one obtains --- after integrating out the heavy fields --- an action of the form~\cite{Gwyn:2012mw}:
\be
\begin{split}
S_{\rm EFT} & = M_{\rm Pl}^2 \int \! d^3xdt  \, a^3 \epsilon H^2 \bigg[   \dot \pi \bigg( 1 - \frac{a^2  \Lambda_{\rm UV}^2}{ \nabla^2} \bigg) \dot \pi -   \frac{(\nabla \pi)^2}{a^2}  \bigg] \\
& +  M_{\rm Pl}^2\int \! d^3xdt  \, a^3 \epsilon H^2 \bigg[  \bigg(   \dot \pi^2 - \frac{(\nabla \pi)^2}{a^2}   \bigg) \frac{a^2  \Lambda_{\rm UV}^2}{ \nabla^2} \dot \pi   +   \frac{2 \tilde c_3}{3}   \bigg(  \dot \pi  \frac{a^2 \Lambda_{\rm UV}^2}{ \nabla^2}  \bigg)^2 \dot \pi    \bigg].\label{EFT-new-physics-1} 
\end{split}
\ee
This action continues to describe a single degree of freedom, and therefore its cutoff energy scale is given by the mass $\Lambda_{\rm UV}$ of the heavy degrees of freedom~\cite{Achucarro:2012sm, Achucarro:2012yr,Gwyn:2012mw}. This version of the EFT may be seen as a non-trivial intermediate completion of the previous one shown in eq.~\eqref{Seft-simple}, with Laplacians $\nabla^2$ modifying the scaling of the operators affecting the evolution of perturbations. This scaling allows the EFT to display a smooth transition by remaining weakly coupled as it runs towards the ultraviolet (UV), where new degrees of freedom become operative. The energy range where this scaling becomes manifest is called the new physics regime~\cite{Baumann:2011su}, a regime where linear perturbation theory is characterized by a dispersion relation, in Fourier space, of the form $\omega^2(k) \propto k^4$. Crucially, if curvature perturbations exit the horizon within this regime,\footnote{A good rule of thumb telling us the value of the wavelength $k^{-1}$ at which curvature perturbations exit the horizon is given by the simple condition $\omega^2 \sim H^2$. Therefore, the freezing of the modes may happen during the new physics regime if the dispersion relation is of the form $\omega^2 \propto k^4$ during horizon exit.} then this time the amplitude of the power spectrum, tensor-to-scalar ratio, and $f_{\rm NL}$ parameters are found to be characterized respectively by:
\be
\Delta_{\mathcal{R}} = \frac{2.7}{100} \frac{H^2}{M_{\rm Pl}^2 \epsilon} \sqrt{\frac{\Lambda_{\rm UV}}{H}} , \qquad   r = 7.6 \epsilon  \sqrt{\frac{H}{\Lambda_{\rm UV}} } ,  \qquad f_{\rm NL} \sim \frac{\Lambda_{\rm UV}}{H} . \label{predictions-EFT-new}
\ee
These expressions may be compared with those of eq.~\eqref{predictions-EFT}: they have the same form but with $c_{\rm s}$ replaced by $\sqrt{H / \Lambda_{\rm UV}} $. In particular, the dependence of both $\Delta_{\mathcal{R}}$ and $r$ on $\sqrt{H / \Lambda_{\rm UV}}$ leads to the same equation \eqref{H-r-Delta} determining the Hubble parameter $H$ in terms of observables. 

While it is not surprising that the new mass scale $\Lambda_{\rm UV}$ shows up in the observables, the fact that they lead to the same relation \eqref{H-r-Delta} suggests that $c_{\rm s}$ and $\sqrt{H / \Lambda_{\rm UV}} $ fulfil similar roles at linear perturbation level. Indeed, as we shall see, they both denote the phase velocity of the Goldstone mode at the moment of Hubble freezing in two different limits. As a result, the two EFT parameterizations are degenerate in the sense that they predict the same relations among observables involving the free field theory. On the other hand, one might have expected that self-interactions would break such a degeneracy by implying different non-Gaussian shapes for these models. We will show that this is not the case. A detailed analysis of the non-Gaussian shapes shows that both theories are indistinguishable for any practical purpose.

To judge the relevance of this situation, let us keep in mind that within the effective field theory framework it is of the utmost importance to understand how measurable --- low-energy --- quantities are related to the free parameters of the underlying theory. If one believes that single field canonical slow-roll inflation is only an effective description embedded in a more fundamental theory containing heavy degrees of freedom, then both \eqref{Seft-simple} and \eqref{EFT-new-physics-1} are equally natural parameterizations. This is because the UV physics responsible for the reduction in the speed of sound, parametrized by $M_2$,  may also contain heavy degrees of freedom, parametrized by $\Lambda_{\rm UV}$. Adopting such a perspective, \eqref{predictions-EFT-new} implies that a non-Gaussian signal would provide information about the ratio $\Lambda_{\rm UV}/H$ (instead of $c_{\rm s}^{-2}$), while the recent results by {\sc Bicep2} \cite{Ade:2014xna} would constrain the quantity $\epsilon \sqrt{H/\Lambda_{\rm UV}}$ (instead of $\epsilon c_{\rm s}$).

Let us examine this claim in the context of a well-studied UV inflationary model: D-brane inflation on a GKP background \cite{Giddings:2001yu}. In such a scenario (see e.g. \cite{Kachru:2003sx}), inflation appears because of the motion of a D-brane in a highly warped throat which is smoothed in the infrared (IR) by fluxes, and glued to a compact internal manifold in the UV. The fluxes are responsible for producing a non-trivial warp factor and for stabilizing the closed string moduli of the Calabi-Yau. The motion of the D-brane may be effectively described by the DBI action which contains higher-order kinetic terms resulting in a reduced propagation speed and a reduced sound horizon $\lambda_H=c_{\rm s}/H$ \cite{Silverstein:2003hf,Alishahiha:2004eh,Chen:2004gc,Chen:2005ad}.  These effects are parametrized by the $M_n$ coefficients of \eqref{Seft-simple}. However, as already mentioned, the presence of background fluxes also results in the stabilization of moduli. These massive scalars are parametrized by the $\Lambda_{\rm UV}$ parameter of \eqref{EFT-new-physics-1}. In the case where the length scale $\lambda_M=M^{-1}$ is small compared to the characteristic length of the perturbations, $\lambda_H$, the effect of these scalars is negligible. The action \eqref{EFT-new-physics-1} becomes relevant in the opposite case.

Finally, let us stress that the action \eqref{EFT-new-physics-1} is constructed entirely within the spirit of ref.~\cite{Cheung:2007st}, where several operators were classified according to their compatibility with the symmetry of the low-energy theory. The operators involved in \eqref{EFT-new-physics-1} satisfy this criterion and their physical interpretation is that they parametrize heavy degrees of freedom. Their relevance or not for CMB observations is a model-dependent question just as in the case of other sets of allowed operators like, for example, extrinsic curvature contributions \cite{Cheung:2007st,Bartolo:2010bj,Bartolo:2010im,Anderson:2014mga}, or Galilean operators \cite{Creminelli:2010qf}. In the absence of a unique UV model, the best we can do is, as usual, parametrize our ignorance and constrain it through actual measurements.

The purpose of this article is to analyze the impact of future measurements --- particularly related to non-Gaussianity --- on discriminating between different models of inflation, described by effective field theories with drastically different parameterizations, such as those of eqs.~\eqref{Seft-simple} and~\eqref{EFT-new-physics-1}. We will pay attention to the role of the non-Gaussianity shapes and show that new signatures are generated in the presence of heavy fields but they are degenerate with those of the low-derivative EFT, to a degree that renders the two descriptions indistinguishable from any practical perspective. What is important though is the precise connection of the observables to the dimensionful parameters of the underlying theory, and we will show how this occurs in our parametrization, so that recent results may constrain the scale of heavy physics directly --- see \cite{Assassi:2013gxa} for similar arguments. This will constitute one of our main results.

We have organized our work in the following way: In Section~\ref{sec:comments}, we begin by explaining the dependence of the three-point amplitude on the scale of UV physics by showing that $f_{\rm NL}$ is related to the phase velocity of the Goldstone boson, which interpolates between the two predictions \eqref{predictions-EFT} and \eqref{predictions-EFT-new}, depending on the value of the combination $c_{\rm s}^2\Lambda_{\rm UV}$ relative to $H$. In Section~\ref{sec:bispec}, we calculate the three-point correlators and extract the precise dependence of $f_{\rm NL}$ on the parameters of the underlying intermediate EFT, which we then invert to obtain constraints, using {\sc Planck} and {\sc Bicep2} results. In Section~\ref{sec:degen}, we comment on the degeneracy of three-point functions of the two effective actions, while we conclude in Section~\ref{sec:conc}.

\section{Comments on the non-linearity parameters} \label{sec:comments}

It is well known that models of inflation with a speed of sound $c_{\rm s}$ different from one are characterized by an enhancement of the equilateral shape of non-Gaussianity, with an amplitude of the order of $c_{\rm s}^{-2}$. At perturbation level, the speed of sound is simply the phase velocity at which Goldstone boson modes propagate in the long wavelength limit $k \ll H$, where $k$ is the comoving momentum of a given mode. Such models arise whenever non-trivial interactions modify the kinetic structure of the inflationary adiabatic curvature perturbations, which at low energies are well parametrized by the action \eqref{Seft-simple}. 

However, as argued in the introduction, it is reasonable to expect that the interactions responsible for introducing a speed of sound $c_{\rm s} \neq 1$ may further modify the kinetic structure at short wavelengths. This is precisely the case for models of inflation where heavy fields interact with curvature perturbations~\cite{Tolley:2009fg, Achucarro:2010jv, Achucarro:2010da, Achucarro:2012sm, Achucarro:2012yr, Gwyn:2012mw}. Here, heavy fields may exchange energy with curvature perturbations producing a mixing between adiabatic and isocurvature modes, resulting in a non-trivial modification of their dispersion relations. In what follows we examine the EFT arising from having integrated out heavy fields that interact with curvature perturbations. For detailed discussions on how this EFT is deduced, see refs.~\cite{Achucarro:2012sm, Burgess:2012dz, Gwyn:2012mw, Cespedes:2013rda, Castillo:2013sfa,Noumi:2012vr}. For other discussions concerning the phenomenology of heavy fields during inflation, see refs.~\cite{Jackson:2010cw, Cremonini:2010sv, Jackson:2011qg, Shiu:2011qw, Cespedes:2012hu, Avgoustidis:2012yc, Gao:2012uq,Gao:2013ota,Gao:2013zga, Pi:2012gf, Achucarro:2012fd, Achucarro:2013cva, Achucarro:2014msa, Mizuno:2014jja, Battefeld:2014aea}.

\subsection{The effective action and free field dynamics}
Integrating out a single\footnote{See \cite{Cespedes:2013rda} and the appendix of \cite{Gwyn:2012mw} for a more general case.} heavy degree of freedom, one deduces the low-energy effective action for the adiabatic perturbation. This action reads\cite{Gwyn:2012mw}
\be 
\begin{split}
S_{\rm EFT} =& -M_{\rm Pl}^2 \int d^3xdt  a^3 \dot H \bigg[    \dot \pi \left( 1 +  \Sigma (\tilde\nabla^2)  \right) \dot \pi -   (\tilde \nabla \pi)^2     +     \big[   \dot \pi^2 -  (\tilde \nabla \pi)^2   \big] \Sigma (\tilde\nabla^2) \dot \pi  \\
& -    \frac{2 \tilde{c}_3  }{3} \dot \pi   \Sigma (\tilde\nabla^2) \left( \dot \pi \Sigma (\tilde\nabla^2)\dot \pi \right)  -    \frac{2\tilde{d}_3}{3} \left ( \Sigma (\tilde\nabla^2) \dot \pi \right ) \left ( \Sigma (\tilde\nabla^2) \dot \pi \right )  \left ( \Sigma (\tilde\nabla^2)\dot \pi \right )  \bigg] ,  \label{EFT-new-physics-2}
\end{split}
\ee
where $\tilde \nabla \equiv a^{-1}\nabla$, and where we have defined:
\be \label{Sigma-def}
\tilde{c}_3 \equiv \dfrac{ c_{\rm s}^2}{ (1 - c_{\rm s}^2 ) }\dfrac{M_3^4}{M_2^4},\quad \tilde{d}_3 \equiv \dfrac{ c_{\rm s}^4 }{ (1 - c_{\rm s}^2 )^2   }  \dfrac{M^2_2}{M^3}\tilde{M}_3, \quad \Sigma (\tilde\nabla^2) =  ( 1 - c_{\rm s}^{2} )  \frac{M^{2} c_{\rm s}^{-2} }{M^2 - \tilde \nabla^2}.
\ee
In these expressions $M$ represents a mass scale characterizing the heavy field sector that has been integrated out, while $c_{\rm s}$ represents the speed of sound of the Goldstone boson modes in the long wavelength limit, given by \eqref{sos}. However, as already stressed in the introduction, the mass of the heavy degree of freedom corresponds to the combination $\Lambda_{\rm UV} = M/c_{\rm s}$, which may be much larger than $M$ if the speed of sound remains suppressed. 

It may be seen that both \eqref{Seft-simple} and \eqref{EFT-new-physics-1} correspond to different limits of this action. More precisely, the action of eq.~\eqref{Seft-simple} is recovered in the limit $H \ll M c_{\rm s}$, whereas the action of eq.~\eqref{EFT-new-physics-1} is recovered in the limit $Mc_{\rm s} \ll H \ll M/c_{\rm s}$. In this sense, the action~\eqref{EFT-new-physics-2} may be thought of as an intermediate completion of the action \eqref{Seft-simple} towards the cutoff scale $\Lambda_{\rm UV}$, incorporating the non-trivial effects from heavy fields that cannot be encapsulated by \eqref{Seft-simple} alone. The last interaction term in \eqref{EFT-new-physics-2} arises from a cubic self-interaction of the heavy field with a dimensionful coupling $\tilde{M}_3$, and was not considered in ref.~\cite{Gwyn:2012mw}, since in this case the equation of motion for the heavy field is non-linear. However, such a term can be treated perturbatively in the interaction picture and we will thus include it in the present analysis. By first considering the action to quadratic order, one may derive the linear equation of motion:
\be \label{pi-full-eom}
\ddot\pi + H \left( 1 - 2\frac{\dot \omega}{H \omega} \right) \dot \pi + \omega^2 \pi = 0,
\ee
where  $\omega$ is given by the dispersion relation, deduced from the quadratic part of the action~\eqref{EFT-new-physics-2}, 
\be
\omega(p) = \sqrt{ \frac{M^2 + p^2}{ M^2 c_{\rm s}^{-2} + p^2}} p, \label{full-modified-dispersion}
\ee
with $p = k/a$, where $k$ denotes the comoving momentum.
Assuming that all modes reach the Hubble scale ($\omega(p) \sim H$) in the dispersive regime $M \ll p \ll \Lambda_{\rm UV}$, or equivalently $c_s^2 \Lambda_{\rm UV}  \ll H \ll \Lambda_{\rm UV}$, the equation of motion \eqref{pi-full-eom} simplifies considerably and the solution for the curvature perturbation in the interaction picture is given by \cite{Baumann:2011su}
\be \label{pi-solution}
\mathcal{R}(z) = \frac{\cal A}{k^{3/2}} \left( \frac{\Lambda_{\rm UV}}{H} \right)^{1/4} z^{5/4}H_{5/4}^{(1)}( z ); \quad z = \frac{H}{2 \Lambda_{\rm UV} } k^2 \tau^2, \quad {\cal A} = - 2^{1/4} \frac{H}{( M_{\rm Pl}^2 \epsilon  )^{1/2}} \sqrt{\frac{\pi}{4}} ,
\ee
where $\tau = - ( H a)^{-1}$ is the usual conformal time and $H^{(1)}$ denotes the Hankel function of the first kind. In the far IR limit $k\tau \to 0$ the previous expression reads
\be \label{R^0}
\mathcal{R}^{(0)}(k) \sim - \dfrac{\sqrt{2}\Gamma(5/4)}{\sqrt{\pi}}\dfrac{H}{ ( M_{\rm Pl}^2 \epsilon  )^{1/2}}\left( \dfrac{\Lambda_{\rm UV}}{H} \right)^{1/4}\dfrac{1}{k^{3/2}},
\ee
and the amplitude of the power spectrum in eq.~\eqref{predictions-EFT-new} is then recovered, i.e. $\Delta_{\mathcal{R}}=\frac{k^3}{2\pi^2} | \mathcal{R}^{(0)}(k)|^2$.
\subsection{The bispectrum amplitude}
In order to understand what the three-point function amplitude probes, it is instructive to see how the operator $\Sigma(p^2)$, defined in \eqref{Sigma-def}, appears in the action. We will only consider momenta within the domain of validity of the effective field theory $p < \Lambda_{\rm UV}$, where the dispersion relation \eqref{full-modified-dispersion} may be approximated by
\be
\omega(p) = \sqrt{\Sigma^{-1}(p^2)} p, \label{full-modified-dispersion-2}
\ee
omitting factors of $(1-c_{\rm s}^2)$. Let us now organize the cubic part of the Lagrangian \eqref{EFT-new-physics-2} using the following notation: 
\bea 
 \mathcal O_{I}^{(3)}  &=& \dot \pi^2 \Sigma (\tilde\nabla^2) \dot \pi,  \label{operator-notation-1} \\ 
 \mathcal O_{II}^{(3)} &=&  \tilde{c}_3 \dot \pi   \Sigma (\tilde\nabla^2) \left( \dot \pi \Sigma (\tilde\nabla^2)\dot \pi \right), \label{operator-notation-2}  \\ 
 \mathcal O_{III}^{(3)}  & = & \tilde{d}_{3} \left( \Sigma (\tilde\nabla^2) \dot \pi \right) \left( \Sigma (\tilde\nabla^2) \dot \pi \right) \left( \Sigma (\tilde\nabla^2)\dot \pi \right), \label{operator-notation-3}  \\  
 \mathcal O_{II'}^{(3)} &=& (\tilde \nabla \pi)^2 \Sigma (\tilde\nabla^2) \dot \pi. \label{operator-notation-4} 
\eea
Since we are interested in computing quantities around the freezing regime when all modes satisfy the horizon crossing condition $\omega(p_*) \sim H$, we are allowed to make the following replacements in these operators: $\partial_t \to \omega(p_*) = H$ and $p^2 \to p_*^2 = H^2 \Sigma_*$, where $\Sigma_*\equiv \Sigma(p_*^2)$.\footnote{Note that this is not a recursive definition, as $p_*$ is determined uniquely by the condition $\omega = H$ and \eqref{full-modified-dispersion-2}, and $\Sigma_*$ is a function of this $p_*$.} Rewriting the kinetic part of the Lagrangian \eqref{EFT-new-physics-2} in terms of $\Sigma_*$, we obtain
\be 
\mathcal O^{(2)}\big|_{\omega = H} =  H^2 \Sigma_* \pi^2, \label{gaussian-eft}
\ee
while the cubic operators may be written as
\bea 
\mathcal O_{I}^{(3)}\big|_{\omega = H} &=& H^2 \Sigma_* \pi^2 \mathcal{R}, \label{op-not-1-fr} \\ 
\mathcal O_{II}^{(3)}\big|_{\omega = H} &=& \tilde{c}_3 H^2 \Sigma_*^2 \pi^2 \mathcal{R},  \\
\mathcal O_{III}^{(3)}\big|_{\omega = H} &=& \tilde{d}_{3} H^2 \Sigma_*^3 \pi^2  \mathcal{R},  \\ 
\mathcal O_{II'}^{(3)}\big|_{\omega = H} &=&  H^2 \Sigma_*^2 \pi^2 \mathcal{R}. \label{non-gaussian-eft}
\eea
From \eqref{gaussian-eft} and \eqref{op-not-1-fr}-\eqref{non-gaussian-eft}, we see that the operator $\Sigma$ appears in the action in the same way that the coupling $M_2^4$ appears in the low derivative EFT \eqref{Seft-simple}, correlating --- via symmetry --- a low phase velocity with a large non-Gaussianity. We thus expect that the value of $\Sigma$ at the Hubble scale determines the amplitude of the three-point function. Indeed, taking the ratio of these expressions with \eqref{gaussian-eft}, we immediately see that the $\Sigma$, $\Sigma^2$ and $\Sigma^3$ operators lead to 
\be \label{all-fnls-Sigma}
f_{\rm NL}^{I} = 1, \quad f_{\rm NL}^{II} = \tilde{c}_3 \Sigma_*, \quad f_{\rm NL}^{III} = \tilde{d}_3 \Sigma_*^2, \quad \text{and} \quad f_{\rm NL}^{II'} = \Sigma_*,
\ee
up to numerical factors that we will include later. To further clarify this result, let us define a phase velocity from \eqref{full-modified-dispersion-2} as
\be 
v_{\rm ph}(p) = \sqrt{\Sigma^{-1}(p^2)}. \label{phase-v}
\ee
The non-linearity parameters \eqref{all-fnls-Sigma} may thus be written as 
\be \label{fnl_full}
f_{\rm NL}^{I} = 1, \qquad f_{\rm NL}^{II} = \frac{ \tilde{c}_3 }{v_{\rm ph}^2(p_*)}, \qquad f_{\rm NL}^{III} = \frac{ \tilde{d}_3 }{v_{\rm ph}^4(p_*)}, \qquad f_{\rm NL}^{II'} = \frac{ 1 }{v_{\rm ph}^2(p_*)}.
\ee
We may now use these relations to obtain a general expression for the amplitude of the three-point functions corresponding to these operators, for the full range of momenta $0 < p < \Lambda_{\rm UV}$. These expressions will depend on the ratio $H/(c_{\rm s}^2 \Lambda_{\rm UV})$ since the dispersive behaviour of the Goldstone boson at freezing depends on this quantity. The operator $\Sigma$ at the Hubble scale may be obtained using the dispersion relation at $\omega(p_*) = H$, which yields
\be \label{pstar-of-x}
p_*^2(x) = \frac{M^2}{2}\left( \sqrt{1+4x^2}-1 \right), \quad v_{\rm ph}^{-2}\left(p_*(x)\right) = \Sigma_*(x) =  \frac{2 c_{\rm s}^{-2}}{1 + \sqrt{1+4 x^2}}; \qquad x \equiv \frac{H}{c_{\rm s}^2 \Lambda_{\rm UV}}.
\ee
Substituting these expressions into \eqref{all-fnls-Sigma}, we obtain
\be \label{all-fnls-full}
f_{\rm NL}^{II} = \frac{2 \tilde{c}_3 c_{\rm s}^{-2}}{1 + \sqrt{1+4 x^2}}, \quad f_{\rm NL}^{III} = \frac{4 \tilde{d}_3 c_{\rm s}^{-4}}{\left( 1 + \sqrt{1+4 x^2} \right)^2}, \quad f_{\rm NL}^{II'} = \frac{2 c_{\rm s}^{-2}}{1 + \sqrt{1+4 x^2}}.
\ee
Taking the two limits $p_*^2 \ll M^2$ and $p_*^2 \gg M^2  $ (or equivalently $x \ll 1$ and $x \gg 1$), we see that the momentum and the phase velocity \eqref{phase-v} at the Hubble scale and the leading predictions for $f_{\rm NL}$ read\footnote{Recall that in the $M\to\infty$ limit, the coefficient $\tilde{d}_3$ defined in \eqref{Sigma-def} and consequently the non-linearity parameter $f_{\rm NL}^{III}$ vanish.}
\be \label{fnl_lin}
p_* = \frac{H}{c_{\rm s}}, \quad v_{\rm ph} = c_{\rm s}, \quad f_{\rm NL}^{II}  = \frac{ \tilde{c}_3 }{c_{\rm s}^2} ,  \quad f_{\rm NL}^{III} = 0, \quad f_{\rm NL}^{II'}  = \frac{ 1 }{c_{\rm s}^2} ,
\ee
for the case $x \ll 1$, and
\be \label{fnl_non-lin}
\begin{split}
& p_* = \frac{H}{v_{\rm ph}} = \sqrt{H \Lambda_{\rm UV}}, \quad v_{\rm ph}(p_*) = \sqrt{\frac{H}{\Lambda_{\rm UV}}}, \\ \quad f_{\rm NL}^{II} =& \tilde{c}_3 \frac{\Lambda_{\rm UV} }{H} , \qquad f_{\rm NL}^{III} = \tilde{d}_3 \left( \frac{\Lambda_{\rm UV}  }{H} \right)^2, \quad f_{\rm NL}^{II'} = \frac{\Lambda_{\rm UV}  }{H},
\end{split}
\ee
for the case $x \gg 1$. (Recall from \ref{fnl_full} that $f_{\rm NL}^{I}$ is independent of $x$). These expressions are in accordance with the $M\to\infty$ limit in which the EFT \eqref{EFT-new-physics-2} flows to the EFT \eqref{Seft-simple}.

Therefore, the predictions \eqref{predictions-EFT} of the low-derivative EFT \eqref{Seft-simple} generalise to the predictions \eqref{predictions-EFT-new} of the EFT \eqref{EFT-new-physics-2}, upon replacing the speed of sound \eqref{sos} with the phase velocity \eqref{phase-v}. In both cases, the non-linearity parameter $f_{\rm NL}$ equals the inverse phase velocity squared. Depending on the value of the parameter $x \equiv H/(c_{\rm s}^2 \Lambda_{\rm UV})$, this phase velocity is related either to the ratio $M_2^4/(|\dot H|M_{\rm Pl}^2)$, or the ratio of the heavy physics scale to the Hubble scale, namely $\Lambda_{\rm UV}/H$. Moreover, in \cite{Gwyn:2012mw,Baumann:2011su} the symmetry breaking scale $\Lambda_{\rm sb}$ and the strong coupling scale $\Lambda_{\rm sc}$ were computed for the theory \eqref{EFT-new-physics-2}. In further support of our claim, let us point out that the same expressions for $\Lambda_{\rm sb},\Lambda_{\rm sc}$ can be derived by taking the analogous expressions for the EFT \eqref{Seft-simple} --- see e.g. \cite{Cheung:2007st} --- and replacing $c_{\rm s}$ with $v_{\rm ph}$ evaluated at the relevant energies (see Sec.~6.2 of \cite{Sypsas:2014aua} for further details). In \cite{Gwyn:2012mw} we proposed that the process of integrating out heavy physics may be thought of as the insertion of an effective UV \emph{medium} through which the IR mode propagates. We see that $f_{\rm NL}$ encodes the ``optical'' properties of this medium, i.e. its \emph{refractive index}.

\section{Bispectra in the presence of heavy fields} \label{sec:bispec}
Let us now compute the shapes of the bispectra in momentum space, defined as
\be  \nn
\langle \hat{ \mathcal{R} }_{\mathbf k_1} \hat{ \mathcal{R} }_{\mathbf k_2} \hat{ \mathcal{R} }_{\mathbf k_3} \rangle =(2\pi)^3 \delta(\mathbf{k}_1 + \mathbf{k}_2 + \mathbf{k}_3) B(k_1,k_2,k_3) ,
\ee 
corresponding to the cubic operators appearing in eqs.~\eqref{operator-notation-1}-\eqref{operator-notation-4}. These can be computed using the $in-in$ formalism \cite{Maldacena:2002vr,Weinberg:2005vy}, according to which the expectation value of an operator $\hat{O}$ is evaluated using
\be \nn
\langle \hat{O} \rangle = \langle 0| \left[ \bar{\mathcal{T}} \exp \left\{ i\int_{-\infty_-}^0 d\tau'\hat H(\tau') \right\} \right] \hat{O} \left[ \mathcal{T} \exp \left\{ -i\int_{-\infty_+}^0 d\tau'\hat H(\tau') \right\} \right] |0\rangle ,
\ee
with $\mathcal{T},\bar{\mathcal{T}}$ standing for time ordering and anti-ordering respectively, and with $\infty_{\pm} = \infty (1 \pm i \epsilon)$.
Using the Baker-Campbell-Hausdorff formula one can expand the previous expression as
\be
\begin{split}
\langle \hat{O} \rangle(\tau) & = \langle 0| \Bigg\{ \hat{O}(\tau) + i \int_{-\infty}^\tau d\tau_1[\hat H(\tau_1),\hat{O}(\tau)] 
  + \ldots \Bigg\} |0\rangle. \end{split}
\label{in-in-0}
\ee
We will focus on the tree-level corrections consisting of the second term of \eqref{in-in-0}, where the operator under consideration is $\hat{O} = \hat{\mathcal{R}}_{\mathbf k_1} \hat{\mathcal{R}}_{\mathbf k_2} \hat{\mathcal{R}}_{\mathbf k_3}$. The field operator $\hat{\mathcal{R}}$ in Fourier space is defined by
\be \nn
\hat{\mathcal{R}}_{\mathbf k}(\tau) = \mathcal{R}_{\mathbf k}(\tau)\hat a_{\mathbf k} + \mathcal{R}^*_{\mathbf k}(\tau) \hat a^\dag_{-\mathbf k},
\ee
where $\mathcal{R}_{\mathbf k}$ denotes the Fourier mode of the field with wavevector $\mathbf k$, and $\hat a^\dag$, and $\hat a$ stand for the usual creation and annihilation operators obeying the canonical commutation relation:
\be \nn
[\hat a_{\mathbf k},\hat a^\dag_{-\mathbf k'}] = (2\pi)^3 \delta({\mathbf k} + {\mathbf k}').
\ee

From now on, we will focus on the part of the bispectrum $B_{II'}$ induced by the operator $ \mathcal O_{II'}^{(3)} $ of eq.~\eqref{operator-notation-4}, the computation of which we write in some detail, and simply quote the results for the other three operators appearing in eqs.~\eqref{operator-notation-1}-\eqref{operator-notation-3}.
In the dispersive limit $p_*^2 \gg M^2$, where momentum dominates over the mass $M$, the Hamiltonian in momentum space is given by
\be \nn
\hat H_{II'}(\tau) = - \int d^3x \hat L_{II'} = \dfrac{1}{(2\pi)^6}\frac{M_{\rm Pl}^2 \epsilon}{H^2} \dfrac{\Lambda_{\rm UV}^2}{H^2} \int \frac{d^3q_1 d^3q_2 d^3q_3}{\tau^3} \frac{q_1^2 - q_2^2 - q_3^2}{2 q_1^2} \hat{\mathcal{R}}'_{q_1} { \hat{\mathcal{R}} }_{q_2} { \hat{\mathcal{R}} }_{q_3} \delta \left({\mathbf q}\right),
\ee
where ${\mathbf q} = \sum {\mathbf q}_i$, and from \eqref{in-in-0}, the first tree-level correction to the three-point correlator reads
\be \nn
(2 \pi)^3 \delta(\mathbf{k}_1 + \mathbf{k}_2 + \mathbf{k}_3)  B_{II'}(k_1,k_2,k_3)  = -i \int_{-\infty}^{0} d\tau \langle [\hat{ \mathcal{R} }_{\mathbf k_1} \hat{ \mathcal{R} }_{\mathbf k_2} \hat{ \mathcal{R} }_{\mathbf k_3} , \hat H_{II'}(\tau) ] \rangle.
\ee
By expanding the commutator and performing the necessary contractions among the operators, we arrive at the final integral which is
\be \label{3pt}
 B_{II'}(k_1,k_2,k_3) =  2 {\rm Im} \Bigg[ \frac{M_{\rm Pl}^2 \epsilon }{H^2} \frac{\Lambda_{\rm UV}^2}{H^2}\frac{k_1^2 - k_2^2 - k_3^2}{2 k_1^2} \mathcal{R}^{(0)}_{k_1} \mathcal{R}^{(0)}_{k_2} \mathcal{R}^{(0)}_{k_3} \int^{0}_{-\infty} \frac{d\tau}{\tau^3} \mathcal{R}'^*_{k_1} \mathcal{R}^*_{k_2} \mathcal{R}^*_{k_3} + \text{perm} \Bigg],
\ee
with $\mathcal{R}_k$ given by \eqref{pi-solution}.
 
Let us first focus on the integral
$$I_{II'} = \int^{0}_{-\infty} \frac{d\tau}{\tau^3} \mathcal{R}'^*_{k_1} \mathcal{R}^*_{k_2} \mathcal{R}^*_{k_3}.$$
Changing the integration variable from $\tau$ to $z = \frac{1}{2 } v_{\rm ph}^2 k_1^2 \tau^2$ (recall $v_{\rm ph}=\sqrt{H/\Lambda_{\rm UV}}$ from eq.~\eqref{fnl_non-lin}) and using the solution \eqref{pi-solution}, we obtain
\be \nn
I_{II'} = \frac{{\cal A}^3}{k_1^{3/2}}  v_{\rm ph}^{3/2} x_2 x_3 \int^{0}_{\infty} dz z^{9/4} H_{1/4}^{(2)}(z)H_{5/4}^{(2)}(x_2^2 z)H_{5/4}^{(2)}(x_3^2 z),
\ee
where we have introduced the ratios $x_2 = k_2/k_1$ and $x_3 = k_3/k_1$.
Taking an analytic continuation $z \mapsto -iz$, so that $H_\nu^{(2)}(-iz) = \dfrac{2}{\pi} (-i)^{-\nu-1} K_\nu(z)$, with $K_\nu$ the modified Bessel function of the second kind, yields
\be \label{integral_II1}
I_{II'} = \frac{{\cal A}^3}{k_1^{3/2}} v_{\rm ph}^{3/2}  \left( \frac{2}{\pi} \right)^3 e^{i\pi/4}  x_2 x_3 \int_{0}^{\infty} dz z^{9/4} K_{1/4}(z)K_{5/4}(x_2^2 z)K_{5/4}(x_3^2 z).
\ee
We may now substitute \eqref{R^0} and \eqref{integral_II1} into \eqref{3pt} and obtain the three-point correlator for the operator $\mathcal{O}_{II'}$.

\begin{figure}[tbh!!!]
\centering
\includegraphics[scale=0.6]{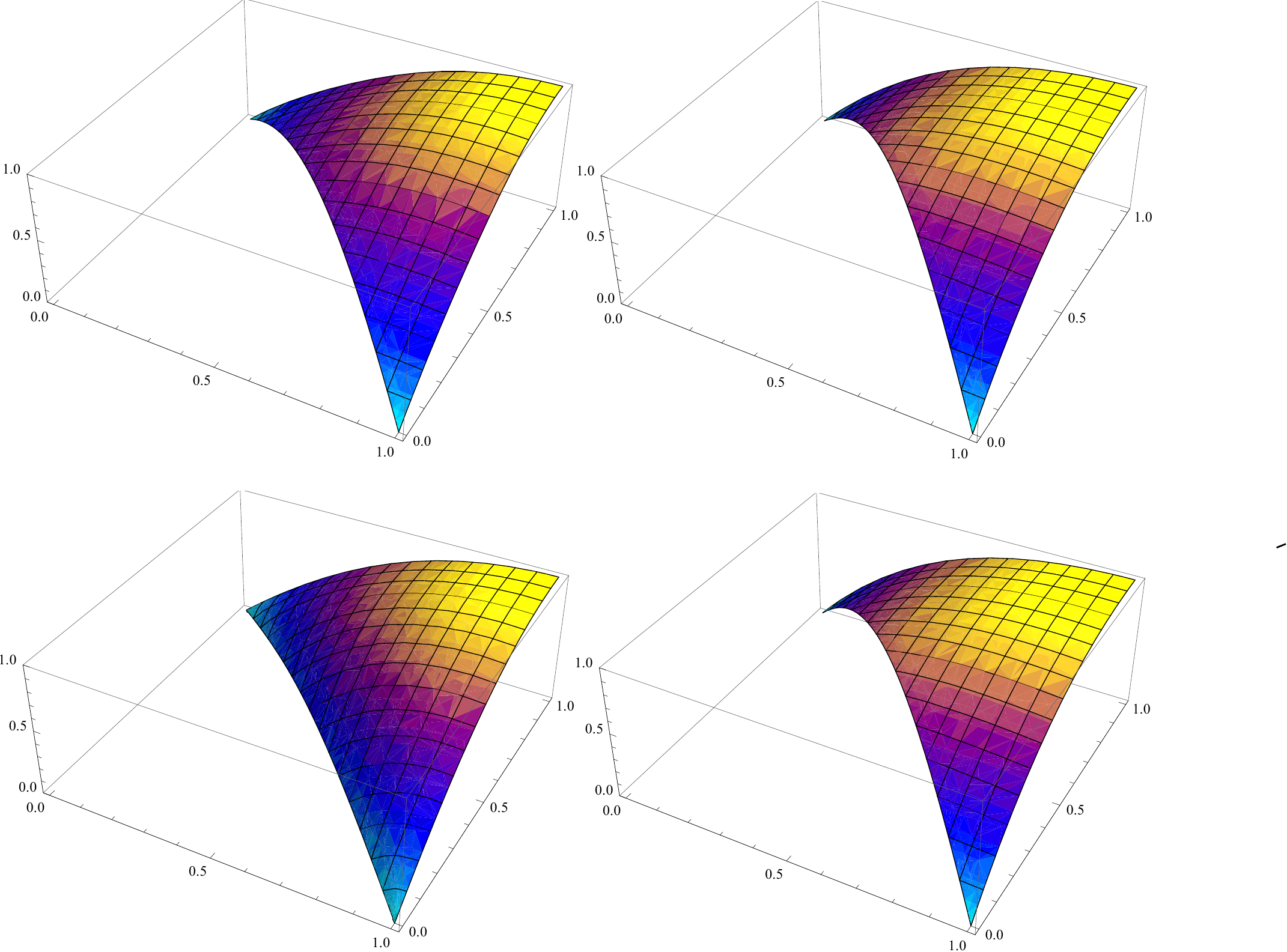}
\caption{\sf The bispectra $x_2^2 x_3^2 S_i(1,x_2,x_3)$ of the effective theory \eqref{EFT-new-physics-2}, normalized to one in the equilateral configuration. Clockwise from top left: $S_{I},S_{II},S_{III},S_{II'}$. $S_{II}$ and $S_{III}$ are highly degenerate but evaluation at the flattened triangle $x_2=x_3=1/2$ reveals their difference.}
\label{fig:shapes}
\end{figure}

In complete analogy, we may derive the expressions for the other operators in eqs.~\eqref{operator-notation-1}-\eqref{operator-notation-3}. Upon defining
\be \nn
f_{\mathrm{NL}}^i = \frac{B_\Phi^i(1,1,1)}{6 k^6 P^2_{\Phi}(k)},
\ee
and using the relation $\Phi= \dfrac{3}{5} \mathcal{R}$,
the three-point functions for the Newtonian potential $\Phi$ read
\be \label{b-shapes}
\begin{split}
  B_{\Phi}^I =  6 P^2_{\Phi}(k) f_{\rm NL}^{I}  S_I^{\rm eq}(1,x_2,x_3),& \quad 
   B_{\Phi}^{II} = 6 P^2_{\Phi}(k) f_{\rm NL}^{II} S_{II}^{\rm eq}(1,x_2,x_3), \\
 B_{\Phi}^{III} =  6 P^2_{\Phi}(k) f_{\rm NL}^{III} S_{III}^{\rm eq}(1,x_2,x_3),& \quad
   B_{\Phi}^{II'} = 6 P^2_{\Phi}(k) f_{\rm NL}^{II'}  S_{II'}^{\rm eq}(1,x_2,x_3),
\end{split}
\ee
where $S^{\rm eq}$ is used to denote the shape function normalized at the equilateral limit $1=x_2=x_3$, and the power spectrum $P_{\Phi}(k)$ is defined by $\langle \mathcal{R}_{\bf k_1} \mathcal{R}_{\bf k_2} \rangle= (2\pi)^3 \delta({\bf k_1} + {\bf k_2}) \frac{25}{9} P_{\Phi}(k)$, and may be computed using the late time solution \eqref{R^0}. The non-linearity parameters read
\be \label{fnls}
\begin{split}
f_{\rm NL}^{I} = \frac{5}{18} \frac{2^{1/4} }{\pi \Gamma[5/4]} \times 0.3549 &, \quad f_{\rm NL}^{II} = \frac{5}{54} \frac{2^{1/4} }{\pi \Gamma[5/4] } \times 0.5369 \tilde{c}_3 v_{\rm ph}^{-2}, \\ f_{\rm NL}^{III} = \frac{5}{36} \frac{2^{1/4} }{\pi \Gamma[5/4]}  \times 0.4999 \tilde{d}_3 v_{\rm ph}^{-4} &, \quad f_{\rm NL}^{II'} = - \frac{5}{72} \frac{2^{1/4} }{\pi \Gamma[5/4]}  \times 7.9071 v_{\rm ph}^{-2 },
\end{split}
\ee
with the phase velocity $v_{\rm ph}$ written in eq.~\eqref{fnl_non-lin}.
The shape functions $S_i$ are given by
\be \label{Ss}
\begin{split}
S_I(1,x_2,x_3) &= \frac{x_2^2 +  x_3^2 + x_2^2 x_3^2}{\sqrt{x_2 x_3}}  \int_{0}^{\infty} dz z^{5/4+2} K_{1/4}(z)K_{1/4}(x_2^2 z)K_{1/4}(x_3^2 z), \\
S_{II}(1,x_2,x_3) &= \frac{1 + x_2^2 + x_3^2}{\sqrt{x_2 x_3}} \int_{0}^{\infty} dz z^{5/4+1} K_{1/4}(z)K_{1/4}(x_2^2 z)K_{1/4}(x_3^2 z), \\
S_{III}(1,x_2,x_3) &= \frac{1}{\sqrt{x_2 x_3}}  \int_{0}^{\infty} dz z^{5/4} K_{1/4}(z)K_{1/4}(x_2^2 z)K_{1/4}(x_3^2 z), \\
S_{II'}(1,x_2,x_3) &= \frac{1 - x_2^2 - x_3^2}{\sqrt{x_2 x_3}} \int_{0}^{\infty} dz z^{5/4+1} K_{1/4}(z)K_{5/4}(x_2^2 z)K_{5/4}(x_3^2 z) + \text{2 perm}, 
\end{split}
\ee
and they are depicted in Fig.~\ref{fig:shapes}. Orthogonal and flattened $(x_2=x_3=1/2)$ shapes can be obtained from linear combinations of the three-point contributions $B_\Phi^i$ in eq.~\eqref{b-shapes} with appropriate values of $\tilde{c}_3$ and $\tilde{d}_3$. For example, the combination $B_{II}+B_{II'}$ with $\tilde{c}_3 \sim 100$ reproduces the orthogonal shape, while with $\tilde{c}_3 \sim 10$ it peaks for the flattened triangle. The same shapes can be obtained for similar values of $\tilde{d}_3$ by combining $B_{III}$ and $B_{II'}$.

In order to make contact with observation, it is necessary to project our predictions onto the templates actually used by experiments. Following \cite{Babich:2004gb} and defining an inner product between two shapes $S_i(1,x_2,x_3)$ and $S_j(1,x_2,x_3)$ as
\be \nn
S_i(1,x_2,x_3) * S_j(1,x_2,x_3) = \int dx_2 dx_3 (x_2x_3)^4 S_i(1,x_2,x_3) S_j(1,x_2,x_3),
\ee
the projected non-linearity parameters can be computed using \cite{Senatore:2009gt}
\be \nn
\left( \begin{array}{c} f_{\rm NL}^{\rm equil}(v_{\rm ph},\tilde{c}_3,\tilde{d}_3) \\ f_{\rm NL}^{\rm ortho}(v_{\rm ph},\tilde{c}_3,\tilde{d}_3) \\ f_{\rm NL}^{\rm flat}(v_{\rm ph},\tilde{c}_3,\tilde{d}_3) \end{array} \right) = \left( \begin{array}{cccc} \frac{S_{I} * S_{\rm equil}}{S_{\rm equil}*S_{\rm equil}} & \frac{S_{II} * S_{\rm equil}}{S_{\rm equil}*S_{\rm equil}} & \frac{S_{III} * S_{\rm equil}}{S_{\rm equil}*S_{\rm equil}} & \frac{S_{II'} * S_{\rm equil}}{S_{\rm equil}*S_{\rm equil}} \\ \frac{S_{I} * S_{\rm ortho}}{S_{\rm ortho}*S_{\rm ortho}} & \frac{S_{II} * S_{\rm ortho}}{S_{\rm ortho}*S_{\rm ortho}} & \frac{S_{III} * S_{\rm ortho}}{S_{\rm ortho}*S_{\rm ortho}} & \frac{S_{II'} * S_{\rm ortho}}{S_{\rm ortho}*S_{\rm ortho}} \\ \frac{S_{I} * S_{\rm flat}}{S_{\rm flat}*S_{\rm flat}} & \frac{S_{II} * S_{\rm flat}}{S_{\rm flat}*S_{\rm flat}} & \frac{S_{III} * S_{\rm flat}}{S_{\rm flat}*S_{\rm flat}} & \frac{S_{II'} * S_{\rm flat}}{S_{\rm flat}*S_{\rm flat}} \end{array} \right)  \left( \begin{array}{c} f_{\rm NL}^{I} \\ f_{\rm NL}^{II} \\ f_{\rm NL}^{III} \\ f_{\rm NL}^{II'} \end{array} \right).
\ee
Using the templates \cite{Creminelli:2005hu,Senatore:2009gt,Meerburg:2009ys,Ade:2013ydc}
\be \nn
\begin{split}
S_{\rm equil}(x_1,x_2,x_3) & = 6 \left( -\frac{1}{x_1^3x_2^3} -\frac{1}{x_1^3x_3^3} -\frac{1}{x_2^3x_3^3} - \frac{2}{x_1^2x_2^2x_3^2} + 
\left[ \frac{1}{x_1x_2^2x_3^3} + 5 \;\rm{perm} \right] \right), \\
S_{\rm ortho}(x_1,x_2,x_3) & = 6  \left( -\frac{3}{x_1^3x_2^3} -\frac{3}{x_1^3x_3^3} -\frac{3}{x_2^3x_3^3} - \frac{8}{x_1^2x_2^2x_3^2} + 3 \left[ \frac{1}{x_1x_2^2x_3^3} + 5 \;\rm{perm} \right]  \right), \\
S_{\rm flat}(x_1,x_2,x_3) & = 6 \left( \frac{1}{x_1^3x_2^3} +\frac{1}{x_1^3x_3^3} + \frac{1}{x_2^3x_3^3} + \frac{3}{x_1^2x_2^2x_3^2} - \left[ \frac{1}{x_1x_2^2x_3^3} + 5 \;\rm{perm} \right]  \right),
\end{split}
\ee
we obtain
\be
\begin{split}
f_{\rm NL}^{\rm equil}(v_{\rm ph},\tilde{c}_3,\tilde{d}_3) & =  0.0157 + 1.8961 v_{\rm ph}^{-2}  + 0.0128 \tilde{c}_3 v_{\rm ph}^{-2} + 0.0167 \tilde{d}_3 v_{\rm ph}^{-4} , \\
f_{\rm NL}^{\rm ortho}(v_{\rm ph},\tilde{c}_3,\tilde{d}_3) & = 0.0005 + 0.1719 v_{\rm ph}^{-2} - 0.0004 \tilde{c}_3 v_{\rm ph}^{-2} - 0.0003 \tilde{d}_3 v_{\rm ph}^{-4}, \\
f_{\rm NL}^{\rm flat}(v_{\rm ph},\tilde{c}_3,\tilde{d}_3) & = 0.0028 + 0.3182 v_{\rm ph}^{-2} + 0.0024 \tilde{c}_3 v_{\rm ph}^{-2} + 0.0031 \tilde{d}_3 v_{\rm ph}^{-4},
\end{split}
\ee
which can be inverted to yield
\be \label{constraint-eqs}
\begin{split}
\frac{\Lambda_{\rm UV}}{H} & = -0.0009 + 38.4502 f_{\rm NL}^{\rm equil} - 29.577 f_{\rm NL}^{\rm ortho} - 209.997 f_{\rm NL}^{\rm flat}, \\
\tilde{c}_3 \frac{\Lambda_{\rm UV}}{H} & = 3.5240 + 46461.8 f_{\rm NL}^{\rm equil} - 41701.4  f_{\rm NL}^{\rm ortho} - 254330 f_{\rm NL}^{\rm flat}, \\
\tilde{d}_3 \frac{\Lambda_{\rm UV}^2}{H^2} & =  - 3.54037 - 39917.2 f_{\rm NL}^{\rm equil} + 35320.9   f_{\rm NL}^{\rm ortho} + 218778  f_{\rm NL}^{\rm flat} .
\end{split}
\ee

From this form one may proceed to input the {\sc Planck} data \cite{Ade:2013ydc} and derive constraints on the values of the dimensionful parameters of the underlying UV theory responsible for inflation.
However, since the variables are correlated, one should use the covariance matrix to compute the error bars. Since such information is not available, what we can do is to examine if theoretically justified values of the parameters $\{ v_{\rm ph},\tilde{c}_3,\tilde{d}_3 \}$ are within observational bounds.

In \cite{Gwyn:2012mw}, we argued that, naturally, the symmetry breaking and strong coupling scales of the EFT \eqref{EFT-new-physics-2} should be of the order of $\Lambda_{\rm UV}$, which implies, via scaling arguments  (see Sec.~6.5.2 of \cite{Sypsas:2014aua}), $\Lambda_{\rm UV}/H \sim 10^2$. Therefore, upon interpreting the {\sc Bicep2} results~\cite{Ade:2014xna} as fixing $H = 10^{14}$ GeV, we are led to the value  $\Lambda_{\rm UV} \sim \Lambda_{\rm GUT}$, which, interestingly, according to \eqref{constraint-eqs} can be achieved with $f_{\rm NL} = {\cal O}(1)$. Such a number is consistent with a high tensor-to-scalar ratio, provided that the slow-roll parameter $\epsilon$ is in the range $10^{-2} - 10^{-1}$, compatible with the {\sc Planck} bound \cite{Ade:2013uln}.
 
Constraints on $M_2$ and $M_3$ can be derived from the requirement that the mass parameter characterizing the heavy field satisfy $M>H$. From $M = c_{\rm s} \Lambda_{\rm UV}$ and the values for $H$ and $\Lambda_{\rm UV}$ quoted above, we obtain $c_{\rm s} \geqslant 0.01$, which\footnote{Notice that in the recent article~\cite{Baumann:2014cja} a new bound on $c_{\rm s}$ was inferred by observing that the tensor-to-scalar ratio receives logarithmic contributions from the speed of sound $r = 16 \epsilon c_{\rm s} (1 + \epsilon \ln c_{\rm s} + \cdots)$. This result modifies the bounds discussed here (in the event that the value of $r$ turns out to be large) however it does not change our more general conclusions regarding the degeneracy between different classes of inflationary models.} leads to $M_2 \leqslant 10 \Lambda_{\rm GUT} \sim 10 \Lambda_{\rm UV}$. A value of the coupling $M_2$ close to the UV scale is consistent with our claim that the physics responsible for reducing the speed of sound also contains heavy degrees of freedom. Note also that a speed of sound of order ${\cal O}(10^{-2})$ is consistent with the requirements $M>H$ and $c_{\rm s}^2 \Lambda_{\rm UV} < H$, with the latter condition implying horizon exit in the dispersive regime. Upon assuming $f_{\rm NL} = {\cal O}(1)$, it follows that $\tilde{c}_3 = {\cal O}(10^3)$ which implies that the $M_3$ parameter should obey $M_3 \leqslant M_{\rm Pl}$.
An upper bound on $\tilde{M}_3$ cannot be derived due to the specific combination of mass scales appearing in $\tilde{d}_3 = c_{\rm s}^4 \frac{M^2_2}{M^3}\tilde{M}_3$ --- see eq.~\eqref{Sigma-def}. The only information that can be extracted from this parameter is $\frac{M^2_2}{M^3}\tilde{M}_3 \leqslant 10^7$, which follows from $f_{\rm NL} = {\cal O}(1)$ and $\tilde{d}_3 = {\cal O}(10^{-1})$.

Finally, let us emphasize again that all these numbers must be taken with caution, since the {\sc Planck} bounds on non-Gaussianity still leave a fairly large parameter space allowed, while the values for $c_{\rm s}$ and $\epsilon$ used to derive them are reasonable assumptions but not experimental data. Furthermore, note that the speed of sound $c_{\rm s}$, or equivalently the mass scale $M$, cannot be probed through our treatment. In order to determine either of these quantities we would have to relax our $p_* \gg M$ condition, so that the parameter $x \equiv H/(c_{\rm s}^2 \Lambda_{\rm UV})$ used in eq.\eqref{pstar-of-x} would show up in the observables. However, this would render the linear equation of motion \eqref{pi-full-eom} hard to solve analytically, and in this work we have not pursued this direction.

\section{Shape degeneracies} \label{sec:degen}
Even though the momentum dependence of the functions \eqref{Ss} is very different compared to the analogous expressions derived from \eqref{Seft-simple}, the resulting shapes, shown in Fig.~\ref{fig:shapes}, are almost identical for the two cases --- see e.g. \cite{Senatore:2009gt}. Indeed, a desirable feature of the EFT \eqref{EFT-new-physics-2}, would be to generate a new distinguishable signature of non-Gaussianities, but evidently this is not the case. Therefore, it is difficult to distinguish the effects of massive fields on the inflaton perturbations using the three-point correlator. In what follows, we comment on the origin of this degeneracy and we argue that this also holds for higher $n$-point correlators.

To clarify the argument, let us first change variables and rewrite the shape integrals of \eqref{Ss} as
\be \label{Is}
\begin{split}
I_I & = \frac{2}{7} \int_{0}^{\infty} dz_1 \left[ z_1^{1/14} K_{1/4} \left(z_1^{2/7}\right) \right] \left[ z_1^{1/14} K_{1/4} \left( x_2^2 z_1^{2/7} \right) \right] \left[ z_1^{1/14} K_{1/4} \left( x_3^2 z_1^{2/7} \right) \right], \\
I_{II} &= \frac{2}{5} \int_{0}^{\infty} dz_2 \left[ z_2^{1/10} K_{1/4} \left(z_2^{2/5} \right) \right] \left[ z_2^{1/10} K_{1/4} \left( x_2^2 z_2^{2/5} \right) \right] \left[ z_2^{1/10} K_{1/4} \left( x_3^2 z_2^{2/5} \right) \right], \\
I_{III} &= \frac{2}{3} \int_{0}^{\infty} dz_3 \left[ z_3^{1/6} K_{1/4} \left(z_3^{2/3} \right) \right] \left[ z_3^{1/6} K_{1/4} \left( x_2^2 z_3^{2/3} \right) \right] \left[ z_3^{1/6} K_{1/4} \left( x_3^2 z_3^{2/3} \right) \right],  \\
I_{II'} &= 2 \int_{0}^{\infty} dz_4 \left[ z_4^{1/2} K_{1/4} \left(z_4^{2}\right) \right] \left[ z_4^{5/2} K_{5/4} \left( x_2^2 z_4^{2} \right) \right] \left[ z_4^{5/2} K_{5/4} \left( x_3^2 z_4^{2} \right) \right] + \text{2 perm},
\end{split}
\ee
where 
\be 
z_1=z^{7/2},z_2=z^{5/2},z_3=z^{3/2} \quad \text{and} \quad z_4=z^{1/2},
\ee
so that all Bessel functions appear in the form $z^{\alpha \nu} K_\nu \left(x_i^2 z^\alpha \right)$. This combination oscillates fast for a large --- sub-horizon --- argument, while for small --- super-horizon  --- $z$, it acquires a constant value (Hubble freezing)
\be \label{bessel-lim}
\left[ z^{\alpha \nu} K_\nu \left(x_i^2 z^\alpha \right) \right]_{z \to 0} = x_i^{-2\nu},
\ee
implying that the integrals \eqref{Is} are dominated by the horizon crossing time $k_*\tau_*v_{\rm ph}(p_*)=1$, with the phase velocity given in \eqref{fnl_non-lin}. As a result, the approximate dependence of the shapes on the ratios $x_2,x_3$ can be extracted by evaluating each term in the limit $z\to 0$. Doing so, we obtain
\be \label{Ss-x}
\begin{split}
S_I(1,x_2,x_3) &\propto \frac{x_2^2 +  x_3^2 + x_2^2 x_3^2}{x_2 x_3}, \quad S_{II}(1,x_2,x_3) \propto \frac{1 + x_2^2 + x_3^2}{x_2 x_3}, \\
S_{III}(1,x_2,x_3) &\propto \frac{1}{x_2 x_3}, \quad S_{II'}(1,x_2,x_3) \propto \frac{ 1 -2 \left( x_2^2 + x_3^2 +x_2^2 x_3^2 \right) + x_2^4+ x_3^4 }{(x_2 x_3)^3},
\end{split}
\ee
where we have also restored the $x_{2,3}$ factors of \eqref{Ss}. These simplified shape functions reproduce --- once multiplied with the measure factor $x_2^2 x_3^2$ \cite{Babich:2004gb} --- the peak structure of the four shapes of Fig.~\ref{fig:shapes}, namely the purely equilateral peak of $S_{II'}$, and the equilateral/flattened peaks of the rest.

In fact, it is straightforward to find a change of variables $z_n(z)$ for which the resulting integrals admit the form 
\be \label{gen-Is}
I = \int dz_n \prod_i \left[ z_n^{\alpha \nu_i} K_{\nu_i} \left(x_i^2 z_n^\alpha \right) \right] \sim \prod_i x_i^{-2\nu_i},
\ee
for any number $n$ of $\Sigma$ insertions. Observing, for example from \eqref{Ss}, that the change in the integrands \eqref{Is} induced by $\Sigma$ is a factor of $a^2 \propto z^{-1}$, we obtain
\be \label{z-change}
z_n=z^{(9-2n)/2} \quad \text{and} \quad z_n=z^{(3-2n)/2},
\ee
for the $\Sigma^n \dot\pi^3$ and $\Sigma^n \dot\pi (\tilde\nabla\pi)^2$ vertices respectively, where the $\Sigma$ operators may be distributed among the three $\pi$'s.
Hence, the three-point integrals are always dominated by the $z\to 0$ limit.

The effect of operators present at the free field level is to change the order of the Hankel functions, i.e. alter the denominators in \eqref{Ss-x}, while those in the interacting part also alter the respective numerators. However, since we obtain polynomials with positive terms in $x_2,x_3$ and we restrict the domain to $x_i \in [0,1]$, we expect the maximum to be at the equilateral configuration, i.e. at $x_2 = x_3 = 1$. In addition, the profiles of the shapes along the $x_3=1-x_2$ line can be shown to be concave curves centered around $x_2=x_3=1/2$, indicating that the flattened configuration also contributes. The degeneracy of the shapes depicted in Fig.~\ref{fig:shapes} and those of the EFT \eqref{Seft-simple} --- cf. \cite{Senatore:2009gt}, is thus slightly lifted to a degree proportional to the ratio of the flattened over the equilateral peaks but realistically speaking this lift is not of significant observational importance. The only way to obtain a non-equilateral shape is to have a polynomial that contains negative terms like in the case of a $k_i \cdot k_j$ interaction. Nevertheless, the shape $S_{II'}$ in \eqref{Ss-x} doesn't have this property. This is because the numerator of this specific vertex vanishes along the $x_3=1-x_2$ line but for insertions of the type $(k_i \cdot k_j)^n,\; n\geq 2$ --- stemming from vertices with a derivative structure of the form $\partial_{ij}^n$ --- this doesn't happen. Indeed, it is well known \cite{Creminelli:2010qf,Bartolo:2010bj} that such operators produce flattened shapes.

In order to illustrate the argument, let us discuss in some detail the form of $S_I$ and $S_{II'}$ in \eqref{Ss-x}. The form of $S_I$ appears as follows: since we have one $\Sigma$, after symmetrizing and pulling out $k_1$ we get $1 + \frac{1}{x_2^2} + \frac{1}{x_3^2} = \frac{x_2^2 +  x_3^2 + x_2^2 x_3^2}{x_2^2 x_3^2}$. Then, the $\mathcal{R}_k^{(0)}$ pieces appearing in the $in-in$ integrals contribute $(x_2 x_3)^{-3/2}$ and the change of variables from $\tau$ to $z$ another $(x_2 x_3)^{3}$ factor, so we are finally left with $\frac{x_2^2 +  x_3^2 + x_2^2 x_3^2}{\sqrt{x_2 x_3}}$, which is what is written in \eqref{Ss}. The final piece comes from the asymptotics \eqref{gen-Is} as $\frac{1}{\sqrt{x_2 x_3}}$, leading to \eqref{Ss-x}. Further multiplying with the measure factor $x_2^2x_3^2$, we obtain the polynomial $x_2^3 x_3 + x_2 x_3^3 + x_2^3 x_3^3$, which obviously has a maximum at $x_2=x_3=1$, while along $x_3=1-x_2$ it reduces to $x_2 - 3 x_2^2 + 5 x_2^3 - 5 x_2^4 + 3 x_2^5 - x_2^6$, which has a maximum at $x_2=1/2$. Polynomials with the same properties can be obtained for $S_{II,III}.$

Similarly, $S_{II'}$ appears as follows: the presence of $k_2 \cdot k_3$ and one $\Sigma$ yields the combination $1-x_2^2-x_3^2$ and the result of \eqref{Ss-x} is reached via symmetrization of the vertex by weighing each factor with the appropriate $x_2^{-2\nu_1}x_3^{-2\nu_2}$ resulting from \eqref{gen-Is}, i.e. $\frac{1-x_2^2-x_3^2}{x_2^{5/2}x_3^{5/2}} + \frac{x_2^2-1-x_3^2}{x_2^{1/2} x_3^{5/2}} + \frac{x_3^2-x_2^2-1}{x_2^{5/2}x_3^{1/2}}$, and then adding contributions from the $\mathcal{R}_k^{(0)}$ and the change of variables, which together yield $(x_2 x_3)^{-1/2}$. We see that the absence of time derivatives acting on all three Hankel functions, thus not lowering their order $\nu$, results in a higher power of $x_2 x_3$ in the denominator that enhances the flattened peak $x_2=x_3=1/2$ but, as already mentioned, in the case of a single $k_i \cdot k_j$ insertion the numerator happens to vanish for this specific configuration. 

This degenerate structure can be traced back to the perturbative scheme:
since the fields involved in the computation of $n$-point correlators are the interaction picture fields, the integrals depend strongly on the behaviour of the solutions of the free theory. These solutions oscillate inside the Hubble radius and freeze outside. Therefore, the main contribution to the integrals comes from the horizon crossing time regardless of the derivative structure of the vertex. For example, the form of the expressions \eqref{Is} is not affected by the $z$ lowering, i.e. they always admit the form \eqref{gen-Is} for any number of $\Sigma$'s via the change of variables \eqref{z-change}. This enables one to just pull out of the integral the momentum dependence according to the spatial derivative structure of each vertex: the only contribution of $\Sigma(\tilde \nabla^2)$ to each shape is the $k^{-2}$ piece coming from $\nabla^{-2}$ (recall the notation $\tilde\nabla \equiv \nabla/a$). This can be better seen in \eqref{Ss-x} before the symmetrization: by observing the last term in each of the numerators of $S_{I},S_{II}$ and $S_{III}$, we see that each $\Sigma$ insertion removes one power of $x_i^2$, which means that 
$$
\int d\tau {\cal F}\left( \frac{a(\tau)^2}{\nabla_i^2} \right) \to {\cal F}\left( \frac{1}{p_*^2}\frac{1}{x_i^2} \right),
$$
where the function ${\cal F}$ is vertex specific and  $p_*$ represents the physical momentum evaluated at $\omega = H$. 

Such reasoning was essentially used in \cite{Cheung:2007st} to explain why equilateral shapes are generically expected for spatial derivative interactions. By exploiting the properties of the free field theory and the perturbative scheme we see that the flattened shape can also be understood. Furthermore, this is why we are allowed to estimate the non-linearity parameters as in eq.~\eqref{all-fnls-Sigma}. Note also that this argument holds for any higher correlator, although these cases admit a much richer structure, while the change of variables \eqref{z-change} can be generalized to any vertex.

Therefore, there are two possibilities that may lift the shape degeneracy between the two parametrizations:
\begin{itemize}

\item[$\star$] Study the three-point correlators including contributions that are higher order in slow-roll. A non-zero spectral index alters the dynamics of the free theory near the horizon crossing regime and modifies the analysis. For instance, in \cite{Burrage:2011hd} new shapes were identified at higher slow-roll order. Furthermore, in \cite{Babich:2004gb} it was shown that non-Gaussianity in the density field, created by non-linear evolution of modes inside the horizon, leads to non-equilateral shapes, a fact which may again be attributed to the scale dependence of the modes, which is a similar effect to the higher slow-roll corrections. Given that quantities such as $r$ and $n_s$ are already measured with high accuracy, this option might not be so unrealistic from an observational point of view.

\item[$\star$] Compute the trispectrum. The arguments given in the previous discussion to explain the degeneracy of the bispectrum for models displaying different scaling might also be valid for the case of higher correlation functions. However, it is reasonable to expect that certain corners in the space of momenta might offer a breaking in the degeneracy of the models at hand.
\end{itemize}
Both of these alternatives presuppose that future cosmological experiments will be able to accurately resolve higher order correlation functions.

\section{Concluding remarks} \label{sec:conc}

Thinking of inflation as a low energy process embedded in a fundamental UV complete theory is a fruitful idea both for understanding the inflationary dynamics of quantum perturbations and gaining insight into the properties that candidate quantum gravity theories should share. The latter perspective has been strengthened, via the Lyth bound \cite{Lyth:1996im, Baumann:2011ws}, by the recent tensor-to-scalar ratio results favouring large field models.
Given our ignorance of such a complete framework, a convenient way to proceed is to use effective field theory techniques and parametrize UV physics in such a way that, in combination with experimental results, one can obtain as much information as possible about the fundamental theory.

Based on the ideas developed in \cite{Cheung:2007st,Gwyn:2012mw}, we have considered an effective field theory describing the dynamics of primordial curvature perturbations with energies close to the Hubble scale $H$, in which UV physics has been integrated. This effective field theory is characterized by a low speed of sound, a non-linear dispersion relation, and cubic self-interactions displaying a non-trivial scaling, as shown in \eqref{EFT-new-physics-2}.

We argued that this type of EFT should arise naturally in string theoretic models and we showed that it can be thought of as an intermediate completion of \eqref{Seft-simple}, with the mass scale $c_{\rm s}^2 \Lambda_{\rm UV}$ serving as the parameter that smoothly  interpolates between the two effective descriptions. This smooth transition makes the predictions of the theories, given in \eqref{predictions-EFT} and \eqref{predictions-EFT-new}, degenerate in the sense that their functional form is identical: both sets depend on the phase velocity of the Goldstone mode, written in eq.~\eqref{phase-v}. In particular, the non-linearity parameter $f_{\rm NL}$ was shown to be proportional to the refractive index of the vacuum on which the Goldstone mode propagates. In the limit $c_{\rm s}^2 \Lambda_{\rm UV} \gg H$, where \eqref{Seft-simple} is the leading description, this is given by $f_{\rm NL} \sim v_{\rm ph}^{-2} = c_{\rm s}^{-2}$ --- see eq.~\eqref{fnl_lin}, while in the opposite case where \eqref{EFT-new-physics-2} becomes relevant, we obtain $f_{\rm NL} \sim v_{\rm ph}^{-2} = \Lambda_{\rm UV}/H$ --- see eq.~\eqref{fnl_non-lin}. Thus, in the latter parametrization the UV scale shows up in the observables and can be constrained via astrophysical surveys.

Working out the exact dependence of the non-linearity parameters on $\Lambda_{\rm UV}$, by computing the three-point functions of the fluctuations, we were able to demonstrate that the {\sc Bicep2} and {\sc Planck} findings are consistent with a value $\Lambda_{\rm UV} \sim \Lambda_{\rm GUT}$.
Even though a desired result would be a distinguishable non-Gaussian signature, this is not the case: the two effective descriptions, \eqref{Seft-simple} and \eqref{EFT-new-physics-2}, predict almost identical shapes. We traced this degeneracy back to the perturbative scheme in use and proposed possible ways to lift it.

\section*{Acknowledgments} 

We would like to thank Jinn-Ouk Gong and David Seery for illuminating discussions. The work of GAP was supported by a Fondecyt Regular Project number 1130777 and a Conicyt Anillo Project number ACT1122. RG is supported by the European Research Council via the Starting Grant Nr. 256994 ``StringCosmOS".

\bibliographystyle{jhep}
\bibliography{biblio}

\end{document}